\documentclass[twocolumn,amsmath,amssymb,prB,superscriptaddress,showpacs]{revtex4-1}

\usepackage{graphicx}
\usepackage{bm}
\usepackage{color}
\begin{document}
	
	\title{Nontrivial transition of transmission in a highly open quantum point contact in the quantum Hall regime}
	
	\author{Changki Hong}
	\affiliation{Department of Physics, Pusan National University, Busan, 46241, Republic of Korea}
	\author{Jinhong Park}
	\affiliation{Department of Condensed Matter Physics, Weizmann Institute of Science, Rehovot, 76100, Israel}
	\author{Yunchul Chung}
	\email{ycchung@pusan.ac.kr}
	\affiliation{Department of Physics, Pusan National University, Busan, 46241, Republic of Korea}
	\affiliation{Department of Condensed Matter Physics, Weizmann Institute of Science, Rehovot, 76100, Israel}
	\author{Hyungkook Choi}
	\email{hkchoi@jbnu.ac.kr}
	\affiliation{Department of Physics, Research Institute of Physics and Chemistry, Chonbuk National University, Jeonju 54896, Republic of Korea}
	\author{Vladimir Umansky}
	\affiliation{Department of Condensed Matter Physics, Weizmann Institute of Science, Rehovot, 76100, Israel}
	\begin{abstract}
		The transmission through a quantum point contact (QPC) in the quantum Hall regime usually exhibits multiple resonances as a function of gate voltage and high non-linearity in bias. Such  behavior is unpredictable and changes sample by sample. Here, we report observation of sharp transition of the transmission through an open QPC at finite bias which was consistently observed for all the tested QPCs.  It is found that the bias dependence of the transition can be fitted to the Fermi-Dirac distribution function through universal scaling. The fitted temperature matches quite nicely to the electron temperature measured via shot noise thermometry. While the origin of the transition is unclear, we propose a phenomenological model based on our experimental results, which may help to understand such a sharp transition. Similar transitions are observed in the fractional quantum Hall regime and it is found that the temperature of the system can be measured by rescaling the quasiparticle energy with the effective charge ($e^*=e/3$). We believe that the observed phenomena can be exploited as a tool for measuring the electron temperature of the system and for studying the quasiparticle charges of the fractional quantum Hall states.
	\end{abstract}
	\maketitle
	
	A quantum point contact (QPC) \cite{van1988quantized,wharam1988one} is the most essential building block of the quantum devices such as quantum dots \cite{meirav1990single}, electron interferometers \cite{yacoby1994unexpected,ji2003electronic}, etc \cite{venkatachalam2012local,bocquillon2013coherence}. In the quantum Hall (QH) regime \cite{klitzing1980new}, it is used to control tunneling between counter propagating edge states, and it provides a useful tool to measure the fractional charge via the shot noise measurement \cite{picciotto1997direct,kane1997observation,chung2003scattering,bid2009shot}. However, the transmission through a QPC usually exhibits multiple resonances as a function of QPC gate voltages and high non-linearity in bias. Such resonances and non-linear behaviors often hamper to develop ideal quantum devices. 
	
	The resonance peaks observed near pinch-off region resemble those of a quantum dot and are explained by the tunneling through localized states in the QPC gap \cite{jain1988quantum,furusaki1998resonant,martins2013coherent}. The exstence of such localized states in an almost closed QPC can be measured as coulomb diamonds and were used as an electron thermometry by Altimiras and coworkers \cite{altimiras2012chargeless}. However, when the QPC is partially open, the resonant peaks are superposed by many other resonances randomly, hence making it very difficult to study their transport behavior systematically. Such random overlap between resonant peaks were usually regarded as the resonant tunneling through multiple localized states which can exist in a rather open QPC. Hence, not much attention has been paid until now. Here, we report sharp transition of the transmission at finite bias for a highly open QPC, which is observed consistently for all the QPCs we have tested. Moreover, it cannot be explained by overlap of multiple resonant peaks. We found that the shape of the transition step is exactly proportional to the Fermi-Dirac distribution function (not the derivative of the distribution function). Also, all the sharp transition traces can be rescaled into a single transition trace, which fits well to the Fermi-Dirac distribution function. The estimated temperature in the fitting is very close to the electron temperature measured via the shot noise thermometry \cite{spietz2003primary}.
	
	The experiments were conducted with three QPCs, QPC A and B with 75nm width and 150nm gap, and QPC C with 150nm width and 300nm gap. The QPC gaps used in this experiment are the typical QPC gaps used for Fabry-Perot interferometer \cite{choi2015robust} and Mach-Zehnder interferometer \cite{weisz2014an} devices. QPCs with wider gap tend to show less resonance peaks but the same chaotic resonances were observed.
	
	The QPCs were fabricated on two different MBE-grown GaAs/AlGaAs heterojunction wafers (for a fractional QH sample: mobility $\mu=$ 3.0$\times$10$^6$cm$^2$/Vs, electron density $n=$ 1.0$\times$10$^{11}$cm$^{-2}$ with 100nm of 2DEG depth; for an integer QH sample: $\mu=$ 3.2$\times$10$^6$cm$^2$/Vs, $n=$ 2.3$\times$10$^{11}$cm$^{-2}$ with 70nm of 2DEG depth).	The measurements were performed by using two different dilution refrigerators, refrigerator A with an electron temperature of 22mK and refrigerator B with an electron temperature of 55mK. Both electron temperatures were measured by using shot noise thermometry \cite{spietz2003primary}.
	\begin{figure}[h]
		\begin{center}
			\includegraphics[width=1\linewidth]{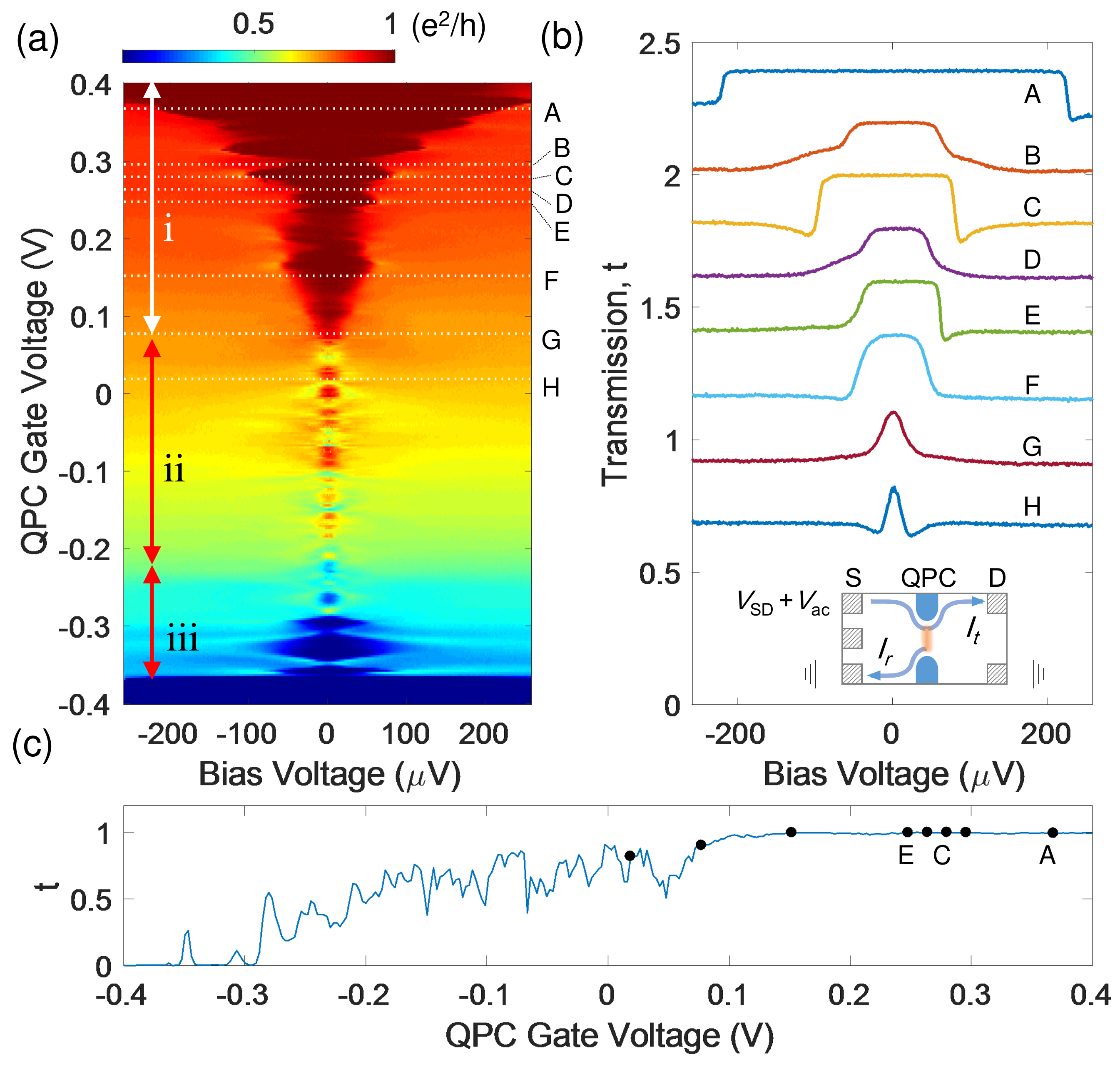}
			\caption{(color online) (a) The transmission through a QPC measured at $\bm{\nu=1}$ (B=9.345T). (b) The transmission of a QPC as a function of bias voltage for various QPC gate voltages. The traces are measured at the white dotted lines in figure (a) and shifted by 0.2 for clarity (bottom to top).  Traces marked as A, C, E show sharp transitions at finite biases. Inset shows the measurement set-up. (c) The transmission measured at zero bias as a function of gate voltage. Points A, C, E marks where the trace A, C, E in figure (b) is measured. Note that the transmission at zero bias is very close to unity.}
			\label{fig:short}
		\end{center}
	\end{figure}
	
	Fig. 1 shows the transmission, $t$ of a QPC A measured as a function of bias and QPC gate voltage at filling factor $\nu=1$ in the refrigerator A (electron temperature of 22mK). The transmission was measured by applying a modulation voltage, $V_{ac}$ of 750kHz, $0.5\mu V_{rms}$ with a bias voltage $V_{SD}$ to the source S and measuring the differential conductance $g=dI_t/dV_{ac}$ at the drain D (Fig. 1(b) inset), where $I_t$ is the transmitted current. The current was measured by monitoring the voltage development over quantum Hall edge using home-made cryogenic low-noise voltage amplifier. The transmission is defined by $g/(\nu e^2/h)$, where $\nu$ is the corresponding filling factor of the quantum Hall states. At zero bias, full transmission ($t=1$) through QPC is observed for the QPC gate voltages above 0.1V (see Fig. 1(a) and (c)). However, the edge state starts to get reflected at finite bias and the transmission drops below 1, as shown in Fig. 1(b). The shape of the peak around zero bias is rather random and slightly asymmetric in bias voltage. The transition slope, $\left|dt/dV_{SD}\right|$ from full to partial transmission varies randomly. Some peaks show slow transitions (trace B, D, F) while others show very sharp transitions (A, C, E). For sharp transitions, we found that the transition does not get infinitely sharper and there is an upper limit for the transition slope. Such sharp transitions were consistently observed for other QPCs we have tested (more than 10, not shown here).
	
	We took the trace A in the Fig. 1(b), which shows the steepest transition slope, and compared with the Fermi-Dirac distribution using Eq.(1), as shown in Fig. 2(a).
	\begin{equation} \label{eq : 1}
F(V_{SD}) = 1-C\frac { 1 }{e^{(eV_0-eV_{SD})/k_BT_0}+1}
\end{equation}
Here, $e$ is the electron charge, $k_B$ is the Boltzmann constant, $V_{SD}$ is the bias voltage. $C$, $T_0$ and $V_0$ are the fitting parameters, which is related to the transition amplitude, the temperature and the bias voltage of the transition center, respectively. The trace was fitted with temperature of 18mK, which is close to the electron temperature of 22mK measured via shot noise thermometry. A similar fit was made with the data measured with the QPC B in a dilution refrigerator B (with an electron temperature of 55mK) as shown in Fig. 2(b). The figure shows almost perfect fit, except for the region at the end of the transition ($V_{SD}>140\mu$V).  Note that using different parameters $C$, $T_0$ and $V_0$ do not give any reasonable fit. After the sharp transition, the transmission usually starts to increase slightly, as it is shown in the Fig. 2.  Most of the transition traces show similar behavior at the end of the transition region, which will be discussed later. In both cases, the electron temperatures measured with shot noise thermometry are slightly higher than the fitted temperatures. However, we believe that such discrepancies are reasonable considering the measurement uncertainty of both techniques.
\begin{figure}[h]
	\begin{center}
		\includegraphics[width=1\linewidth]{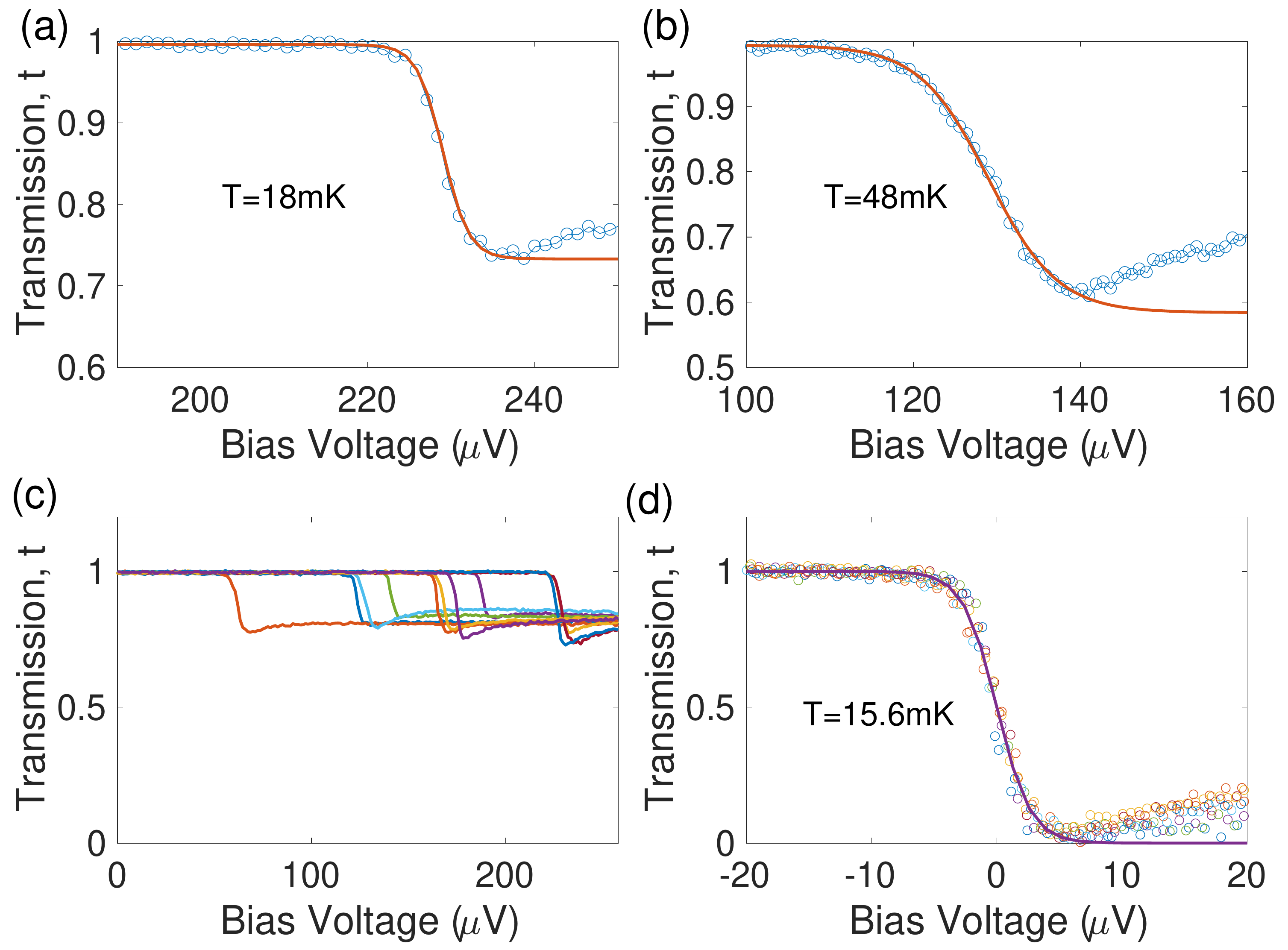}
		\caption{(color online) (a) The trace (circles) showing a sharp transition (trace A in the Fig. 1(b)) is fitted with Fermi-Dirac distribution by using Eq. (1) with the temperature of 18mK (solid line). The electron temperature measured via shot noise thermometry was around 22mK.  (b) Similar fitting but with a trance measured with a QPC B in a refrigerator B (with an electron temperature of 55mK). (c) Sharp transitions (10 traces) measured with QPC A at 3 different magnetic fields (B=9.165T, 9.354T, 9.4T) at $\bm{\nu=1}$. (d) The traces (circles) in (c) were rescaled and shifted to position around zero bias voltage and fitted with Eq. (1) (solid line) with temperature of 15.6mK.}
		\label{fig:short}
	\end{center}
\end{figure}

Fig. 2(c) shows 10 traces of sharp transitions measured with QPC A at three different magnetic fields (in $\nu=1$ plateau) in the refrigerator A. The traces were individually fitted and the average temperature was found to be 15.6mK with the standard deviation of 2.1mK. By shifting and renormalizing the traces using their individual fitting parameters $C$ and $V_0$, the traces were rescaled and plotted in Fig. 2(d). The solid line is the fit using Eq. (1) with the average temperature of 15.6mK.  The figure clearly shows that the transition traces can be universally scaled into a single trace which can be represented with the energy distribution of the Fermi-Dirac function. 

Since the resonant behavior and the non-linear conductance of a QPC under high magnetic fields are usually attributed to the tunneling through localized states \cite{jain1988quantum,furusaki1998resonant,martins2013coherent}, we compare our results with the above model. Fig. 3(a) is a schematic diagram showing the tunneling through localized states in a QPC. Let's assume that a droplet of localized states exists between counter propagating edge states with some finite energy $E_C$ due to charging energy. This allows us to adopt quantum dot analogy except for some subtle differences, which will be discussed later. The tunneling from the transmitting edge T to the reflecting edge R is suppressed at zero bias, due to the charging energy, which allows perfect transmission through the QPC. When the bias is applied larger than the charging energy, then the tunneling from transmitting edge T to the reflecting edge R is allowed via the localized state, hence reduces the transmission.
\begin{figure}[h]
	\begin{center}
		\includegraphics[width=1\linewidth]{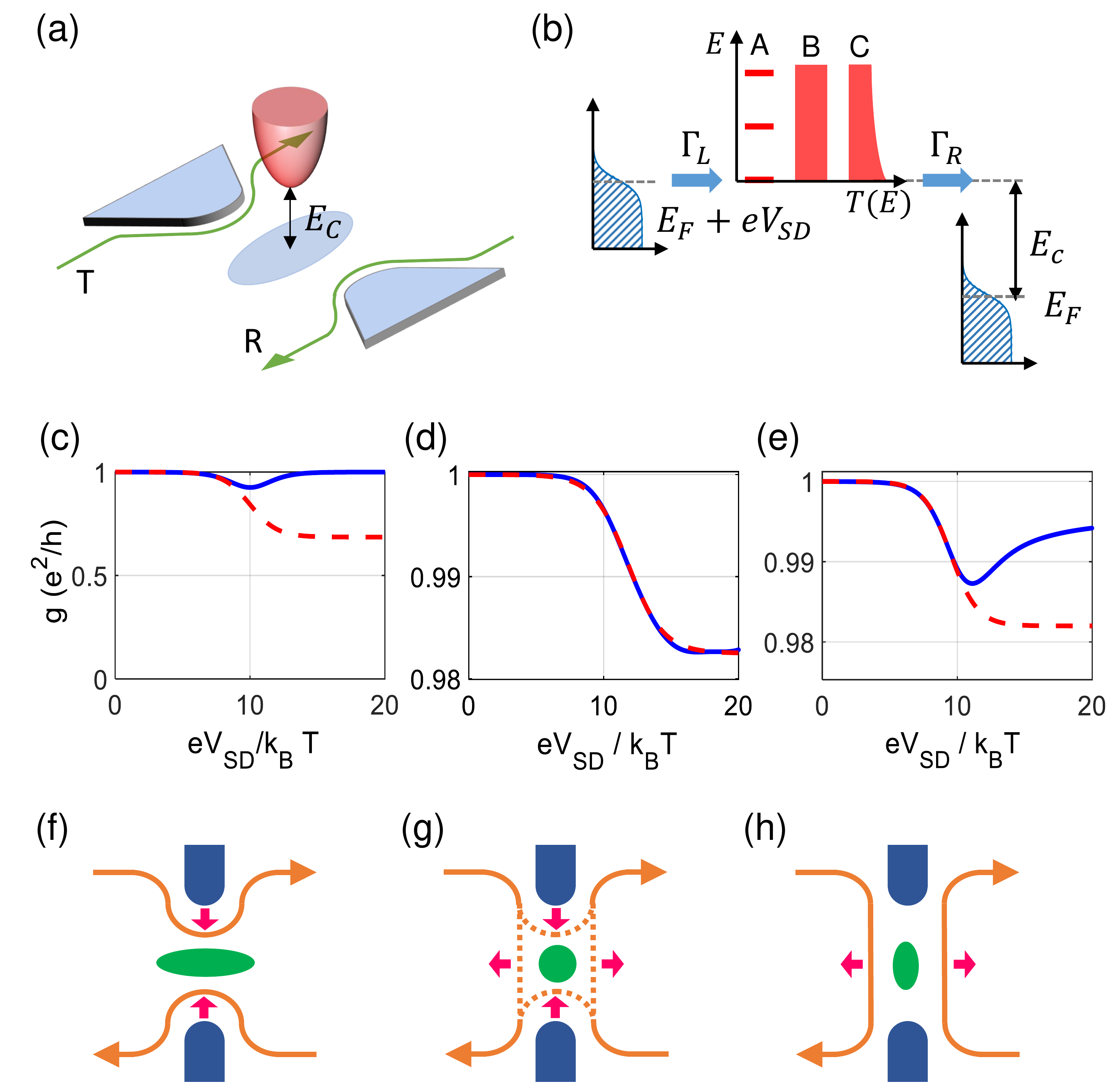}
		\caption{(color online) (a) The localized states (with the charging energy $E_C$) formed between QPC gates, which allow tunneling between the transmitting edge channel T and reflecting edge channel R (b) Left Fermi sea is biased with a voltage $V_{SD}$ and is coupled to the localized state with A discrete, B constant and C energy dependent tunneling density energy state in energy.  The right Fermi sea is grounded. $\Gamma_L,\Gamma_R$  is the tunneling rate between the localized states and the left and the right Fermi sea, respectively. Using a lattice model [14], we calculate the transmissions (denoted as solid lines in (c), (d) and (e)) through the localized states with the three different energy dependencies ((c) 0D, (d) 2D, (e) 1D) of the transmission probability. We used parameters $\hbar\Gamma_L/k_BT=10$ and $E_C/k_BT =10$ for drawing the solid lines of (c), (d), and (e). Solid lines are fitted to Eq. (1) (dashed lines). We used fitting parameters $C=\pi\hbar\Gamma_L/k_BT$, $\frac{eV_0}{k_BT}=10$, $\frac{T_0}{T}=1$ for Fig. 3(c), $C = 0.0175$, $\frac{eV_0}{k_BT}=11.8$, $\frac{T_0}{T}=1.3$ for Fig. 3(d), and  $C = 0.018$, $\frac{eV_0}{k_BT}=9.5$, $\frac{T_0}{T}=1$ for Fig. 3(e). The schematic diagrams showing possible localized states (green island) formed in (f) relatively open, (g) intermediate, (h) relatively closed QPC.}
		\label{fig:short}
	\end{center}
\end{figure}

The sharpness of the transition can be naively explained by the energy level broadening of the localized state. If the energy level broadening in the localized state is much smaller than the temperature broadening ($\hbar\Gamma \ll k_BT$, where $\Gamma = \Gamma_L+\Gamma_R$) then the transition will be sharp and the slope will be only determined by the temperature broadening, whereas, at the opposite limit, the transition will be slow and the slope will be determined by the energy level broadening which is bigger than the temperature broadening. Hence the maximum transition slope is limited by the temperature broadening, which explains why the observed transition does not get infinitely sharper. Up to this point, the quantum dot analogy works perfectly to explain our experimental results. However, the quantum dot analogy fails to explain why the transition reflects the Fermi-Dirac distribution function itself, not the derivative of the distribution function, considering that the transmission is determined by measuring the differential conductance. In general, the differential conductance for a tunneling device can be written as follows \cite{buttiker1992scattering}.
\begin{equation} \label{eq : 2}
g = \frac{ dI_t }{dV_{SD}}=\frac{ e^2 }{h}\int_{-\infty}^{\infty}dE\left(\frac{-\partial f(E-eV_{SD})}{\partial E}\right)T(E)
\end{equation}
Here, $h$ is the plank constant, $f(E)$ is the Fermi-Dirac distribution function and $T(E)$ is the transmission probability. It can be easily seen from the equation that the differential conductance will be proportional to $\frac{\partial f(E-eV_{SD})}{\partial E}|_{E=E_C}$ when the transmission probability is given by a delta function (which is the case for an ideal quantum dot with discrete energy states) while it will be proportional to $f(E_C-eV_{SD})$ when the transmission probability has a sudden jump from zero to a finite value at $E_C$ (which corresponds to the case of an electron transport through a two-dimensional electron puddle).

Fig. 3(c), (d) and (e) are the calculated transmissions through the localized states with the three different energy dependencies of the transmission probability shown in Fig. 3(b). We theoretically consider a tight binding model of a lattice $1\times 1$ (Fig. 3(c)), $40\times 40$ (Fig. 3(d)), and $100\times 1$ (Fig. 3(e)) with hopping $t_0$ to describe various localized states in a QPC. We attach the edge channels T and R to the sites (1, 1) (Fig. 3(c)), (20, 1) and (20, 40) (Fig. 3(d)), and (33, 1) and (66, 1) (Fig.3(e)) with coupling strengths $\Gamma_L=\Gamma_R$. We used parameters $\hbar\Gamma_L/k_BT$= 0.1 and $E_C/k_BT$=10 for drawing (c), (d), and (e), and parameters $t_0/k_BT$=2 for (d) and $t_0/k_BT$=10 for (e).	As it can be seen from the figures, the calculated transmission through the one-dimensional lattice shown in Fig. 3(e) is quite consistent with the observed experimental results. A monotonic increase in the transmission after the sharp transition shown in the experiments may be attributed to the van Hove singularity of finite size 1D wire. We speculate that the localized states might be formed in a highly open QPC because the local filling factor in the QPC area is effectively lowered by the QPC potential \cite{hashisaka2015shot}. This lowered filling factor may lead to formation of an unwanted electron puddle in the QPC area by electrostatic Coulomb interaction \cite{chklovskii1992electrostatics} and the composite edges between the QPC area and vacuum \cite{rosenow2010signatures}.

In Fig. 1(a), the width of the Coulomb diamond in bias voltage is shrinking in the region i, as the QPC closes. The result can be explained by considering the charging energy between the edges and the localized states. When QPC start to pinch, the tunneling is taking place mainly between the upper and the lower edges through the localized states in the gap of a QPC, as shown in Fig. 3(f). As the QPC closes more, the capacitance between the edges and the localized states becomes larger (because the distances between them become closer), hence reducing the charging energy, which reduces the width of the Coulomb diamonds. In the strongly pinched regime, where electrons tunnel from the left edge to the right edge as shown in Fig. 3(h), both edges are pushed away, as the QPC closes. This increases the charging energy hence making Coulomb diamond wider, which can be seen in the region iii of the Fig. 1(a).

In the intermediate regime (region ii in Fig. 1(a)), the localized states experience the capacitance from all the surrounding edges, as shown in Fig. 3(g). The widths of Coulomb diamonds are roughly the same, regardless of QPC gate voltages. In this regime, the capacitance cannot be determined solely by the geometrical distance between the edges and the localized states because the QPC gate voltage also modifies the quasiparticle populations in the edge states, which also alters the capacitance \cite{16}. As the QPC closes, the upper and lower edges are getting closer to the localized states and depopulated while it is the opposite for the left and right edges. Such counteraction may keep the overall capacitance stay roughly the same, hence keeping the Coulomb diamond widths constant. Here, the quasiparticle can tunnel to any edge states, which might be responsible for the chaotic resonances observed. Overall, the tunneling through localized states (with a finite charging energy) roughly explains observed non-linearity and resonances in all QPC voltage ranges, including the sharp transition observed in a highly open quantum point contact.
\begin{figure}[h]
	\begin{center}
		\includegraphics[width=1\linewidth]{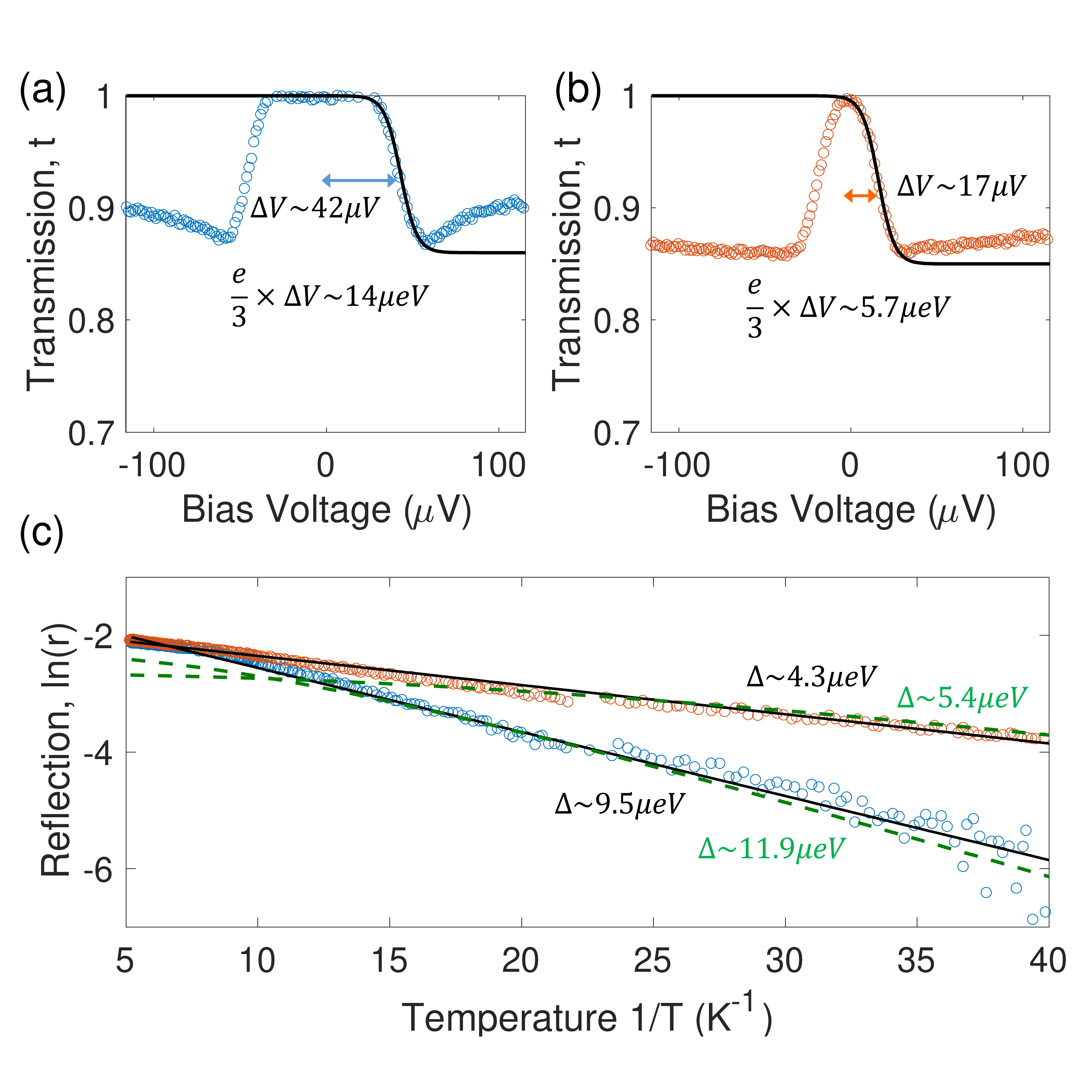}
		\caption{(color online) Two sharp transition traces observed at $\nu=1/3$ showing energy gap $14\mu$eV (blue circles) in (a) and $5.7\mu$eV (red circles) in (b) respectively. Both are fitted with Fermi-Dirac distribution by using Eq. (1) with the temperature of 18mK (solid line). (c) The reflections measured at zero bias. Blue circles are corresponding to the trace of (a) and the red ones to that of (b). The black solid lines are the fitting for equation $r\propto e^{-\Delta/k_BT}$ and the green dashed lines are the fitting for equation $r\propto \cosh^{-2}(\Delta/2.5k_BT)$ (details are in the main text).}
		\label{fig:short}
	\end{center}
\end{figure}

Similar sharp transitions were also observed for the $\nu = 1/3$ fractional quantum Hall state. Two sharp transition traces were shown in Fig. 4(a) and (b). To fit the data with a temperature which is reasonably close to the electron temperature of the refrigerator A (22mK), the electron charge $e$ in Eq. (1) has to be replaced to the effective charge $e^*=e/3$, which is the quasiparticle charge of $\nu = 1/3$ fractional quantum Hall state. With the replacement, both traces were fitted well with the temperature of 18mK, which is close to the temperature measured with shot noise thermometry (22mK). It calls for further theoretical investigation because the $\nu = 1/3$ fractional quantum Hall edge state is believed to be described by the chiral Luttinger liquid theory, where the state does not follow the Fermi-Dirac distribution function \cite{luttinger1963exactly,haldane1981luttinger,wen1990chiral}.

We measured the reflected $r$ at zero bias as a function of a temperature at the same fractional filling factor $\nu=1/3$. We estimate the energy gap $\Delta$ by two different equations $r\propto e^{-\Delta/k_BT}$ and $r\propto \cosh^{-2}(\Delta/2.5k_BT)$. The first equation assumes simple thermal excitation and the second equation assumes tunneling through thermally broadened continuous energy levels \cite{altimiras2012chargeless,beenakker1990theory}. The blue and red circles in Fig. 4(c) are the reflections measured at zero bias of the traces in Fig. 4(a) and (b) and the corresponding energy gaps are 9.5$\mu$eV (11.9$\mu$eV) and 4.3$\mu$eV (5.4$\mu$eV) by using the first (second) equation, respectively. The fitting was more reasonable with the first equation, as it can be seen from the figure.	The energy gap can be also obtained from the transition traces shown in Fig. 4(a) and (b) by extracting the fitting parameter $V_0$ from Eq. (1), which roughly corresponds to the bias voltage of the transition mid-point. Again, multiplying the effective charge $e/3$ rather than $e$ to the bias voltage gives comparable energy gaps measured with the thermal activation experiment, which are $14\mu$eV and $5.7\mu$eV for Fig. 4(a) and (b), respectively. The energy gaps measured with thermal excitation experiment(with the first equation) were slightly smaller ($\sim$ 3/4) than the values estimated from the transition traces. Note that the temperature in Fig. 4 (c) is the lattice temperature measured with a thermometer in the refrigerator and is generally slightly lower than the electron temperature, which may results in energy gap to be smaller. Thus, the results are consistent with the tunneling through localized states with a finite charging energy gap.

The above analysis shows that the quasiparticle charge of a fractional quantum Hall state can be estimated by fitting the non-linear conductance transition, which agrees well with the thermal activation measurement in a highly open QPC. These measurement techniques can be used to confirm the reported shot noise measurement results, for example the observation of super Poissonian noise in fractional quatum Hall regime \cite{rodriguez2002super} and the observation of e/3 charge through a local fractional quantum Hall state in a QPC at $\nu=1$ quantum Hall state \cite{hashisaka2015shot}.

To summarize, we have observed a sharp transition of QPC transmission at finite bias in a highly open QPC in the quantum Hall regime. The transition traces in bias voltage fit almost perfectly with Fermi-Dirac distribution function. Also, it was found that all the sharp transition traces can be rescaled into a single transition trace, which fits to the Fermi-Dirac distribution function. Similar transition is observed in a fractional quantum Hall regime and the temperature of the system is measured by rescaling the quasiparticle energy with the effective charge ($e^*=e/3$). We believe that the observed phenomena can be exploited as a handy tool for measuring the electron temperature of the system.
\newline	

We thank M. Heiblum for discussion and D. Mahalu, N. Ofek, E. Weisz, I. Sivan, R. Sabo, and I. Gurman for their help in experiment. This work was partially supported by the National Research Foundation of Korea grant (NRF-2014R1A2A1A11053072).

\begin{thebibliography}{29}
\bibitem [{16()}]{16}%
	\BibitemOpen
	\href@noop {} {}\bibinfo {note} {The capacitance between two plates is given
		by $C=q/V$. The change in charge population $q$ on the plate without the
		voltage difference change between the plates can be regarded as the change in
		effective capacitance. This is not possible for ordinary capacitor with fixed
		distance between two metal plates. However, in our experiment, the voltage
		difference between the edge state and the localized state is fixed by the
		bias voltage, while the quasiparticle population in the edge state is changed
		by the QPC gate voltage. Hence the quasiparticle population change in the
		edge state can be regarded as the change in the effective capacitance between
		the edge state and the localized state.}\BibitemShut {Stop}%
\end{thebibliography}
\end{document}